\documentclass{article}
\usepackage{graphicx} 
\usepackage{subcaption}
\usepackage{amsmath} 
\usepackage{xcolor}

\title{Connecting strain rate dependence of fcc metals to dislocation avalanche signatures}
\author{M. Aissaoui, C. Kahloun, O.U. Salman, S. Queyreau}
\date{Dec. 2025}

\begin{document}

\maketitle

\begin{abstract}
    Strain rate sensitivity is a key feature of material deformation, whose importance is growing both because miniaturized components experience higher effective rates and because small‑scale simulations increasingly probe such conditions. As a dynamical characteristic, strain rate dependence is shown to be intimately connected to dislocation avalanches, which are a fundamental mechanism of dislocation dynamics. Using carefully designed, state‑of‑the‑art dislocation dynamics simulations in the intermediate range $\dot{\epsilon} \in 5-1000$ s$^{-1}$, we show that increasing strain rate promotes the activation of a growing number of stronger sites. The dislocation microstructure progressively rearranges into configurations with shorter segments. Dislocation avalanches become larger through the superposition of simultaneous events and because stronger obstacles are required to arrest them. As a result, the avalanche statistics are strongly affected by strain rate, with a reduced power‑law regime and an increasing power‑law exponent. Larger avalanches, in turn, lead to an enhanced dislocation storage rate. Contribution from collinear systems to avalanches and cross‑slip activity decreases, altering the fraction of screw dislocations and the resulting microstructure. These results provide an original mesoscopic picture of rate sensitivity in this strain‑rate range and offer a mechanistic interpretation of existing observations and findings from experiments and simulations.
\end{abstract}

\section{Introduction}

Strain rate sensitivity is a key feature of material deformation in industrial applications, where components experience varied or non‑monotonic rates during processing and service. Miniaturization of components is also synonymous with higher effective deformation rates. Understanding strain rate dependence has therefore become of utmost importance, especially since many simulation techniques, such as Molecular Dynamics \cite{bulatov2023one,fan2021strain} and Discrete Dislocation Dynamics (DDD) \cite{fan2021strain,sparks2018nontrivial}, must operate at high strain rates to remain numerically efficient; this raises the question of how to rationalize and compare results obtained at very different strain rates.

At very low, \emph{conventional} strain rates and moderate temperatures, deformation is controlled by thermally activated unpinning of dislocations from obstacles such as junctions or solutes. The activation rate for unpinning follows an Arrhenius law, so to sustain a higher strain rate the system must increase the resolved shear stress to reduce the activation barrier, leading to the well‑known weak logarithmic increase of flow stress with strain rate in fcc metals \cite{follansbee1985mechanical,klepaczko1986rate,regazzoni1987dislocation}. At very high strain rates, say above $\approx 1000$ s$^{-1}$ for Cu, dislocation motion becomes limited by phonon drag, the flow stress rises steeply with imposed strain rate, and a non‑Arrhenius, often nearly linear, scaling with strain rate emerges, consistent with Orowan’s law \cite{regazzoni1987dislocation,fan2021strain}. The transition between these two regimes of strain rates remains insufficiently understood and must be thoroughly investigated in bulk systems before any attempt at quantitative modeling. 

Experimental observations on Cu single crystals deformed at different strain rates \cite{edington1969influence} revealed an increase in dislocation density and a marked change in microstructure, with activation of secondary slip systems and an increase in screw dislocation content as strain rate rises. The latter suggests a possible modification of dynamic recovery processes. Similar features have been reported following strain rate changes \cite{takeuchi1976variation}, where discontinuous slip activity appears on the primary slip system after a rate jump. The strain rate dependence of Frank–Read sources has also been examined, with their activation stress becoming rate sensitive at large imposed strain rates \cite{gurrutxaga2015mechanisms}, although such sources are not the dominant mechanism of dislocation multiplication in pure systems. These experimental features therefore call for a more complete mechanistic interpretation.

Strain rate dependence is inherently a dynamical property of the dislocation microstructure. Most existing studies interpret it in terms of single‑dislocation mechanisms, yet dislocation dynamics in crystals proceeds through collective motion in the form of avalanche‑like plastic bursts \cite{richeton2005breakdown,weiss2015mild}, observed at both the micro \cite{csikor2007dislocation} and mesoscale \cite{devincre2010scale,aissaoui2025physical}. It thus becomes clear that the strain rate dependence of plastic deformation must be intimately linked to the properties of dislocation avalanches, a connection that has yet to be fully elucidated.

The size of dislocation avalanches typically follows a power‑law distribution $P(\Delta \gamma)\propto \Delta \gamma^{\alpha}$, whose exponent may characterize the universality class of the underlying dislocation mechanisms. However, this exponent is found to depend strongly on the imposed strain rate, as inferred from strain increments on deformation curves in pillar experiments \cite{friedman2012statistics,sparks2018nontrivial,sparks2019avalanche} and in 2D and 3D DDD simulations \cite{papanikolaou2012quasi,kurunczi2021dislocation}. In pillars, exponents increase from about -2 to -1.3 as avalanches increasingly overlap, although the finite pillar size competes with the intrinsic dislocation length scales. Contradictory results exist: in \cite{kurunczi2023avalanches}, a constant exponent $\alpha = -1$ was reported in bulk DDD simulations at large strain rates, corresponding to the phonon‑drag‑controlled regime, and these exponents differ from those measured from the discrete part of acoustic emission in macroscale experiments at very low strain rates \cite{weiss2015mild}.

Here, we propose a unifying mesoscale picture for the intermediate regime where the material response evolves from Arrhenius to non‑Arrhenius strain ‑rate ‑controlled behavior. We show that the strain rate dependence of the plastic response of fcc metals is intimately related to the kinetics of dislocation avalanches. To this end, state‑of‑the‑art DDD simulations are employed, incorporating two key rate‑dependent mechanisms: (i) phonon‑drag‑controlled dislocation mobility and (ii) thermally activated cross‑slip of screw dislocations. The resulting deformation curves display higher flow stresses and larger multiplication peaks as strain rate increases, reflecting enhanced dislocation interactions and storage. It is demonstrated that, to sustain higher strain rates, stronger configurations associated with shorter segments are activated simultaneously and in increasing numbers. Consequently, avalanches become larger, all primary systems contribute almost equally, and the relative contribution of cross‑slip systems decreases. Power‑law exponents $\alpha$ then evolve from values characteristic of macroscale experiments at low strain rates toward values closer to those of the phonon‑drag‑controlled regime. The strain rate is finally shown to strongly affect collinear systems and cross‑slip activity, thereby explaining the associated changes in microstructure formation.

\section{Methodology}

We conducted a comprehensive study using large scale DDD simulations with the $microMegas$ code, which has been described in details in 
\cite{devincre2011modeling, queyreau2020dislocation} and making use of the Cai et al.'s non-singular elastic theory for dislocation interactions \cite{cai2006non,arsenlis2007enabling,queyreau2014analytical}. Simulation conditions were carefully tailored to provide meaningful avalanche statistics data \cite{aissaoui2025physical}. This section only covers the specific conditions for the simulations carried out here.

We decided to focus on a tensile loading along the $[001]$ direction in order to obtain stable multiple slip condition, that is particularly relevant for comparison with loading conditions observed within grains of polycrystals under large deformation. The shear modulus of copper is taken as $\mu = 42$ GPa and its Poisson’s ratio as $\nu = 0.34 $, with a Burgers vector magnitude for primary perfect dislocations as $b= \frac{a_0}{2}|111|= 2.5 $ \AA, and $a_0$ the lattice parameter. The microstructures are generated from prismatic loops randomly positioned within a simulation box, following the \emph{principle of similitude} for dislocation microstructures. Their initial length is taken as $l_0 = 10 /\sqrt{\rho_0}$, with $\rho_0 = 10^{12}$ m$^2$ the initial dislocation density.The impact of dislocation interactions and dislocation density was shown in a previous study \cite{aissaoui2025physical}. The dimensions of the orthorhombic box are chosen as to follow the similitude principle with $L_B = 10 /\sqrt{\rho_0}$. [001] microstructures are typically associated to four active slip systems out of eight possible ones. Prismatic loops are made of long segments belonging to one of the four slip systems, that will be referred to as primary systems, while shorter segment are set for the other four slip systems, referred to by simplicity as collinear or cross slip system. The active slip systems lead to a stable deformation along [001], and this modeling strategy allows the expected dislocation storage and hardening rate to be recovered.

Periodic boundary conditions are imposed at the faces of the simulation box. The time step is chosen as $dt = 0.1$ ns and we will see that this value is sufficient to capture the fast avalanche kinetics while reproducing the experimental strain rate sensitivity of Cu. Focusing on the athermal regime of plastic flow, the mobility law defining the dislocation segment 'i' glide velocity is taken as $v^i = \tau_{eff}^i b /B(T)$, typical of the phonon drag mechanism, with $\tau_{eff}^i$ the effective stress felt by the segment. The viscous drag coefficient, is set to $B = 5.5 \cdot 10^{-5} \ Pa \cdot s^{-1}$ in agreement with experimental evaluations in copper at room temperature. 

Our mesoscopic simulations include two rate dependent mechanisms: i) the dislocation glide associated to the phonon drag mechanism and ii) the thermally activated cross-slip mechanism for screw dislocations.  The main series of simulations was carried out on Cu single crystals under constant applied strain rate $\dot{\epsilon _a}$ spanning almost three orders of magnitude from 5 to 1000 s$^{-1}$, with and without cross-slip enabled. These values of strain rate $\dot{\epsilon _a}$ along with the large initial dislocation density, correspond clearly to the regime of the plastic flow controlled by forest hardening as shown by \cite{fan2021strain}. The largest strain rate is just below the strain rate controlled regime (see supplementary regime). We will see that the plastic behavior is still strongly impacted by the strain rate. Our simulations correspond to closed loop control that model loading control used in macroscale experiments, for which the applied stress is updated as $d\sigma = E \left(\dot\epsilon_a - <\dot \epsilon_p>\right)dt$, where $E$ is the young modulus. (This relation illustrates the possible links existing between mechanical behavior, the imposed strain rate and avalanches kinetics through $<\dot \epsilon_p>$ ). The time average of the simulated plastic deformation rate $\epsilon_p$ is carried over a fixed window to smooth the plastic bursts characteristic of avalanches. We made sure that the window size do not impact avalanche statistics \cite{aissa2025}. 

We also systematically modified  the following parameters that are expected to impact strain rate effects: viscous coefficient $B$  from $5 \cdot 10^{-6}$ to $5 \cdot 10^{-4}\ Pa \cdot s^{-1}$, time step $dt$ from $0.01$ to $0.1$ ns, and initial dislocation density $\rho_0$ from 5\, 10$^{10}$ to 2\, 10$^{13}$ m$^{-2}$. These parameters impact the quantitative effects discuss here but not the qualitative strain rate effect. For the sake of clarity and conciseness, we will only refer to these simulations when needed.


Even with sufficient data, analyzing avalanche statistics can be a difficult task. Here, we reemploy the simulation design from our previous study \cite{aissaoui2025physical} and built upon the careful methodology derived by Clauset and coworkers \cite{clauset2009power}. Avalanche features and statistics are defined from the sampling of the deformation curves as typically done in the literature. Our careful modeling design lead to great amount of avalanche data between 10$^3$ to 10$^5$ plastic events per simulation, depending on the strain rate. To provide a quantitative description of the avalanche statistics from the simulations, most of our data is fitted using the robust maximum likelyhood methodology \cite{clauset2009power} by a truncated power law with an exponential decay. The fits are based on data truncated beyond the lower $\Delta \gamma_{min}$ cutoff for the power law regime.  Curves (dashed lines) and fit parameters are generally shown in figures.

\section{Results}

\subsection{Strain rate sensitivity of the plastic behaviour}

\begin{figure}[h!]
    \centering
    \begin{subfigure}[b]{0.45\textwidth}
        \centering
        \includegraphics[width=\textwidth]{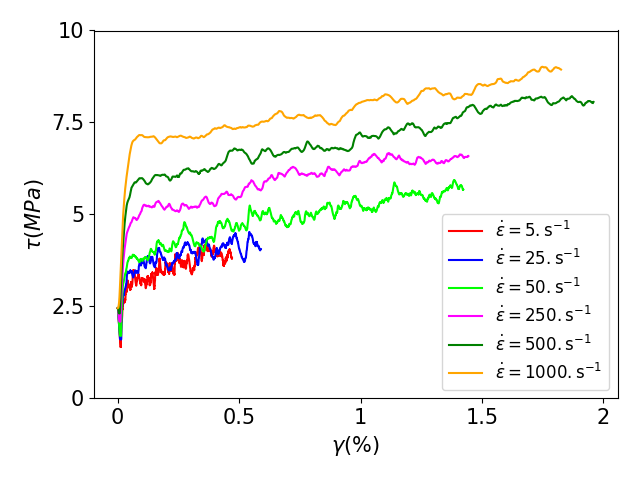}
        \caption{}
    \end{subfigure}
    \hfill
    \begin{subfigure}[b]{0.45\textwidth}
        \centering
        \includegraphics[width=\textwidth]{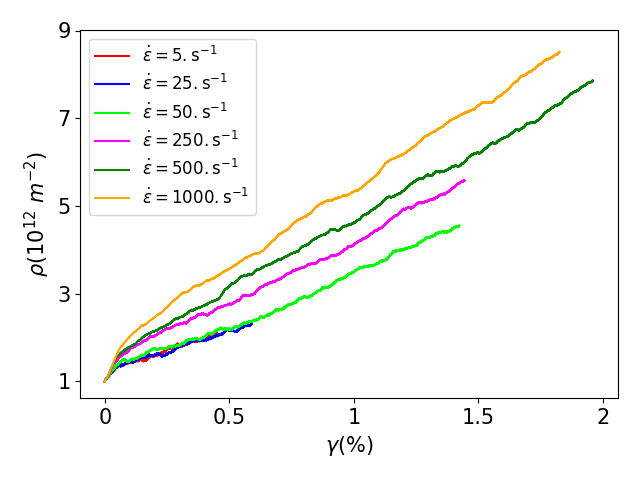}
        \caption{}
    \end{subfigure}
    \vfill
    
    \begin{subfigure}[b]{0.45\textwidth}
        \centering
        \includegraphics[width=\textwidth]{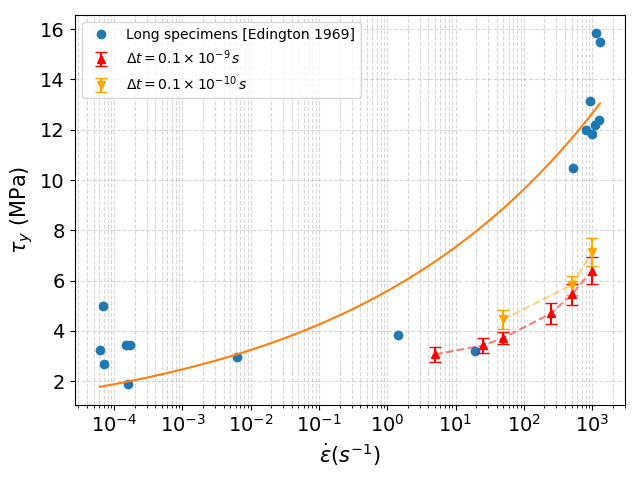}
        \caption{}
    \end{subfigure}
    \caption{impact of the strain rate on the plastic deformation at the mesoscale as simulated by DDD simulations. (a) Typical evolution of the resolved shear stress $\tau$ as a function of the total shear strain $\gamma$ for increasing strain rates $\dot{\epsilon}$ for [001] single crystals. (b) Corresponding evolutions of the dislocation density $\rho$ as a function of the strain $\gamma$. (c) Evolution of the flow stress $\tau_y$ with imposed strain rate and comparison with experimental data from \cite{edington1969influence}.  
    }
\label{fig:epsi-point1}
\end{figure}

We first illustrate that the plastic behaviour of Cu at the mesoscale show a remarkable sensitivity to the imposed strain rate, which is in a qualitative and quantitative agreement with the literature on single crystals. Figure \ref{fig:epsi-point1}(a) shows the typical deformation curves obtained for various strain rates. As expected, the flow stress $\tau_y$ increases with the strain rate. The simulated stress–strain curves exhibit serrations that originate from the intermittent and collective motion of dislocations, commonly referred to as dislocation avalanches. The serrations seem to diminish in number and intensity as strain rate increases. This will be discussed in details further in the paper.


Although the flow stress depends on the strain rate, the strain hardening rate, defined as $\theta = \frac{d\tau}{d\gamma}$, remains approximately the same across all simulations. The hardening rates are also close to the experimental values of $\theta_{[001]} \approx \mu/150$ obtained for copper single crystals loaded along the $[001]$ axis \cite{takeuchi1976variation} 
For the largest applied strain rates $\dot \epsilon > 250$ /s, a multiplication peak seems to be present at the onset of the plastic flow, and this will be confirmed a bit later. In these instances, the nominal [001] hardening rate is thus recovered for the second half of the deformation.

Figure~\ref{fig:epsi-point1}(b) shows now the evolution of dislocations density $\rho$ as a function of the shear strain $\gamma$. Interestingly, the dislocation density increases significantly when increasing the strain rate. This constitutes the second key mesoscopic feature associated to the increase of the strain rate. An increased dislocation density was also obtained from TEM on Cu single crystals oriented for single slip and similar strain rate range \cite{edington1969influence}. The increased dislocation storage is barely noticeable for low strain rates below $\dot{\epsilon}_a < 50$ s$^{-1}$.

 To be more quantitative, the strain rate sensitivity is illustrated in Figure \ref{fig:epsi-point1}(c), where $\tau_y$ is plotted as a function of the strain rate $\dot{\epsilon}$. The evolution follows a non linear relationship sometimes reported as $ \tau_y = A \dot{\epsilon}^m$, with $m$ the so-called strain rate sensitivity parameter  $m = \frac{\partial \ln \tau_y}{\partial \ln \dot{\epsilon}}\Big|_{T,\epsilon} $ at constant temperature $T$ and deformation $\epsilon$. Our data are compared with flow stress measured in deformed Cu single crystals oriented for single slip conditions by Edington et al. \cite{edington1969influence}. Our data is in good qualitative and quantitative agreements with these experimental results. {To match the experimental and our simulation data,  the strain rate sensitivity parameter $m$ has to be vary with strain rate. From our data, $1/m$ decreases from $\approx 20$ at $\dot \epsilon$ = 5 /s to $1/m \approx 7$ at $\dot \epsilon$ = 5 /s. The flow stress variations for large strain rates is however lower than those seen in experiments. It can be argued that strain rate sensitivity is anisotropic and depends on the orientation \cite{takeuchi1976variation}, with more active slip systems, [001] orientation is expected to be less strain rate sensitive. For low strain rates in copper single crystals, $1/m \to 230$ making Cu rather insensitive to strain rate in that range \cite{edington1969influence,takeuchi1976variation}. For large strain rates, $m$ may tend to 1 in the strain rate hardening regime (see supplementary material).}


\subsection{Classical crystal plasticity framework}

For the range of dislocation density and imposed strain rate considered here, we can show that the simulated plastic flow is controlled by the forest mechanism and not strain rate hardening regime where phonon dragging is the controlling mechanism \cite{edington1969influence} following the analysis proposed by Fan et al \cite{fan2021strain} (see Supplementary materials)

\begin{figure}[h]
    \centering
        \includegraphics[width=0.45\textwidth]{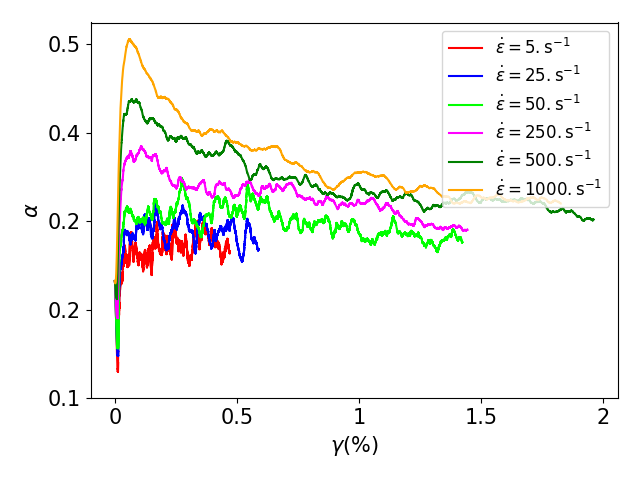}
    \caption{Evolution of the average interaction coefficient $\bar \alpha(\dot\epsilon)$ during deformation and impact of the strain rate.}
    \label{fig:beta-coefficient} 
\end{figure}

Following the classical and physically based modeling of crystal plasticity, the critical stress is expressed in its scalar form for sake of simplicity as \cite{kocks2003physics, devincre2008dislocation}: 
\begin{equation}
    \tau_c = \alpha \mu b \sqrt{\Sigma_1^{n_{sys}} \rho_j },
    \label{eq:tau_c}
\end{equation}
where $\rho_j$ is the 'j' slip system density among all systems $n_{sys}$. Note that the $\tau_c$ may sometimes be further decomposed into a quasistatic term and a strain rate dependent term. But, we use equation \ref{eq:tau_c} above as a way to interpret our results and not a modeling effort yet. $\bar\alpha$ is the so-called interaction coefficient measuring the average strength of dislocation interactions for the considered [001] orientations. Our data showed that $\tau_c(\dot{\epsilon})$ is a function of $\dot{\epsilon}$ and that the dislocation density is increased. Assuming that this equation still holds, we can follow the evolution of the interaction coefficient $\bar\alpha(\dot{\epsilon}) =  \tau_c(\dot{\epsilon}) /\mu b \sqrt{\sum \rho_j}$ with the strain rate as seen in the figure \ref{fig:beta-coefficient}. The average coefficient $\bar\alpha$ clearly increases with the strain rate. A peak on the evolution can be seen on the largest strain rate $\dot \epsilon > 250$ /s, confirming the presence of multiplication peaks seen on the deformation curves in figure \ref{fig:epsi-point1}.a). This means that, as strain rate increases, the average strength of activated configurations increases as will be confirmed further. Please note that $\bar\alpha$ decreases with the dislocation density as the equation above neglects the logarithmic term in the self-stress expression \cite{devincre2006physical,queyreau2009slip}. When computing the evolution of the interaction coefficient corrected for this effect, the obtained curve shows the same qualitative hierarchy between the different curves as function of strain rate, the gap between the low and large strain rate curves is however increased as the latter are associated to larger dislocation densities.

A second key ingredient in the crystal plasticity framework has to do with the dislocation density evolution on active slip systems  'i' \cite{kocks2003physics}:
\begin{equation}
    \frac{d\rho^i}{d\gamma^i} = \frac{1}{b} \left( \frac{1}{L_i} - y\rho^i \right),
    \label{eq:storage}
\end{equation} 
where the first term in the rhs corresponds to the dislocation storage. $L_i$ is the so-called  mean free path, which depends on the interactions between dislocations, the applied stresses, and the crystal orientation, and loading history \cite{queyreau2023saturation}. The second term in the rhs of the equation represents the effect of dynamic recovery with parameter $y$ proportional to the critical annihilation distance for screw dislocations \cite{kubin2013dislocations}.  At small strain such as here, the dynamic recovery term can be neglected, which yields to:
$
\frac{d\rho_i}{d\gamma_i} = \frac{1}{bL_i} = \frac{\tau_c}{\mu b^2 K_{hkl}} = \frac{\beta \sqrt{\rho}}{b K_{hkl}}.
$
The slope representing this evolution was estimated using linear regression, and a dimensionless mean free path coefficient $K_{hkl}$ is evaluated. Devincre, Kubin, and Hoc showed that the storage mechanism of dislocation is correlated with avalanches of dislocations, as dislocation storage occurs at the end of a plastic burst \cite{devincre2008dislocation}. We will see that avalanche sizes increase with the strain rate. We computed the coefficient $K_{hkl}$ for the various strain rate, and past the multiplication peaks when values are identical across the all simulations and close to those reported by Devincre, Kubin, and Hoc \cite{kubin2008modeling} (see table \ref{tab:Khkl_muA_strain_rate}). This means that the storage mechanism are virtually identical, which was suggested in \cite{edington1969influence}. The increased storage is therefore  solely the result from the strongest configurations activated associated to larger coefficient $\alpha(\dot{\epsilon})$.

We solved Eq. \ref{eq:storage} numerically using the evolution of $\alpha(\dot\epsilon, t)$ seen in the simulation, and this confirms that the additional storage seen as strain rate increases can be described entirely by the CP framework confirming that the storage mechanism is unchanged. 


\begin{table}[h!]
\centering
\begin{tabular}{c c c c}
\hline
Strain rate $\dot{\varepsilon}(s^{-1})$ & $K_{001}$ \\
\hline
5     & 5.87 \\
25    & 6.58 \\
50    & 6.38 \\
250   & 5.96 \\
500   & 6.53 \\
1000  & 5.93 \\
\hline
\end{tabular}
\caption{Mean free path coefficient $K_{001}$ calculated past the multiplication peak as function of the imposed strain rate. Reference value is $K_{001} \approx 5$ in \cite{devincre2008dislocation}.}
\label{tab:Khkl_muA_strain_rate}
\end{table}




\subsection{Distributions of stresses triggering avalanches }

\begin{figure}[ht]
\centering
\begin{minipage}{0.45\linewidth}
    \centering
    \includegraphics[width=\linewidth]{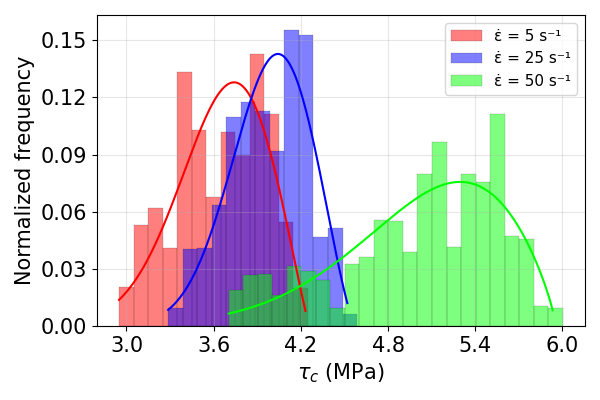}
\end{minipage}
\hfill
\begin{minipage}{0.45\linewidth}
    \centering
    \includegraphics[width=\linewidth]{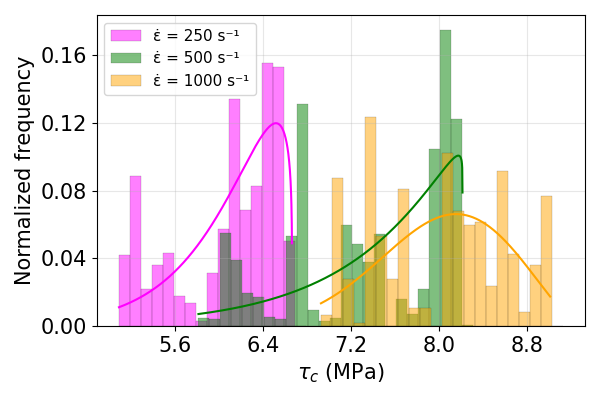}   
\end{minipage}

\vspace{0.3cm}

\begin{minipage}{0.45\linewidth}
    \centering
    \includegraphics[width=\linewidth]{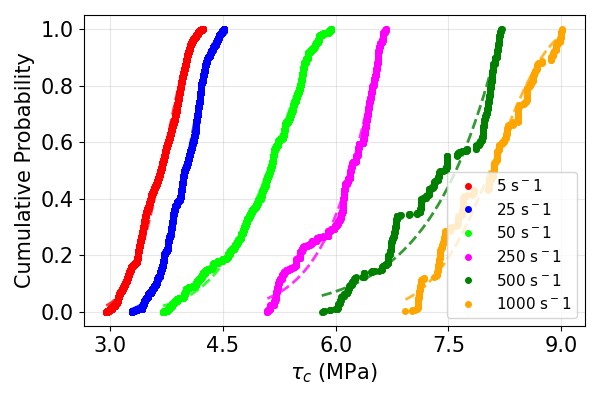}

\end{minipage}
\hfill
\begin{minipage}{0.45\linewidth}
    \centering
    \includegraphics[width=\linewidth]{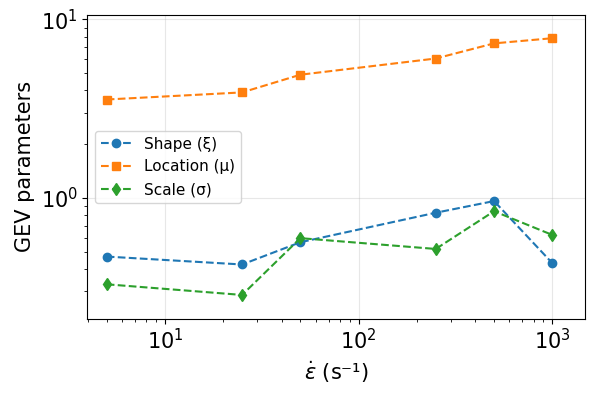}
    
    \label{fig:tauact_histograms}
\end{minipage}

\caption{a-b) Density histograms of activation stresses as function of strain rate. c) Cumulative histograms of the same data. d) Evolution of the parameters of the GEV distributions modeling the simulation data.}
\label{fig:tauact_histograms}
\end{figure}

Next, we show that as strain rate increases, stronger configurations are activated corresponding to a growing portion of the microstructure. In the absence of lattice friction and solutes as here, the stresses triggering avalanches correspond to critical stresses associated to the destruction of junctions or activation of source-like configurations. The stress to activate a single Frank-Read source was shown to be constant and strain rate independent for low strain rate up to a transition strain rate, and eventually increases for larger strain rates \cite{gurrutxaga2015mechanisms}. However, strain rate sensitivity of materials cannot be understood on the sole basis of individual sources of fixed properties as dislocation microstructures dynamically evolve during straining. Here we show, that increasing the strain rate yields to a rearrangement of the dislocation microstructure and the activation of progressively stronger configurations.

For this, we employed the methodology in \cite{aissaoui2025physical} where the deformation curves were analyzed to detect avalanches. At the onset of an avalanche, the applied stress correspond to the activation stress $\tau_{app}=\tau_{act}$ required to activate one or several configurations of the microstructure. The corresponding histograms are typically not always well defined. But thanks to our large simulation data, density histograms are well defined here at least for strain rates between 5-50 /s, as seen in figure \ref{fig:tauact_histograms}. Histograms are lesser defined for larger strain rates, as we will see that there is less data for these simulations. The main feature is that the histograms are progressively shifted towards larger activation stresses as strain rate increases.


In \cite{aissaoui2025physical}, we showed that the histogram of avalanche triggering stresses was consistent with a well defined distribution of the form: 
\begin{equation}
f(\tau_c) =
\frac{k}{\bar{\tau}_c}
\left(\frac{\tau_c}{\bar{\tau}_c}\right)^{k-1}
\exp\!\left[-\left(\frac{\tau_c}{\bar{\tau}_c}\right)^k\right]
\end{equation}
with $\tau_c$ given by equation \ref{eq:tau_c} and $k=0.6 \pm0.05$ uniquely defined for a fixed strain rate ($\dot \epsilon = 50 /s$). To have a little more flexibility in the fitting procedure required for accounting for histogram variations as function of strain rate, we quantify the evolution of the histograms of $\tau_{act}$ here with a generalized extreme value distribution of the form:
\begin{equation}
f(\tau_{\mathrm{act}}) =
\left[
1 + \xi \left( \frac{\tau_{\mathrm{act}} - \mu}{\sigma} \right)
\right]^{-1/\xi}
\end{equation}

The quantitative parameters of the GEV distribution are given in figure \ref{fig:tauact_histograms}.d) where we see a global increase of the three parameters with the strain rate. 


For the sake of the argument, let us assume that the microstructure is associated configurations with activation stresses following a Gaussian distribution. For the symmetrical loading direction such as the [001] axis, let us assume further that the activation configuration and stresses for all slip systems follow the same distribution. The configurations that actually get activated among all contained within the microstructure are the ones corresponding to lowest activation stresses. The distribution of the actually activated configurations is the one that we have access to through the figure \ref{fig:tauact_histograms}. As strain rate increases, configurations associated to larger activation stresses are required to produce sufficient strain rate. Larger activation stresses will correspond to a growing part of the the Gaussian distribution, and more sites will be activated simultaneously on different systems. Conversely, the part of the microstructure participating to plastic deformation increases with the strain rate.

\subsection{Slip systems contribution to avalanches }

\begin{figure}[ht]
\centering
\begin{minipage}{0.48\linewidth}
    \centering
    \includegraphics[width=\linewidth]{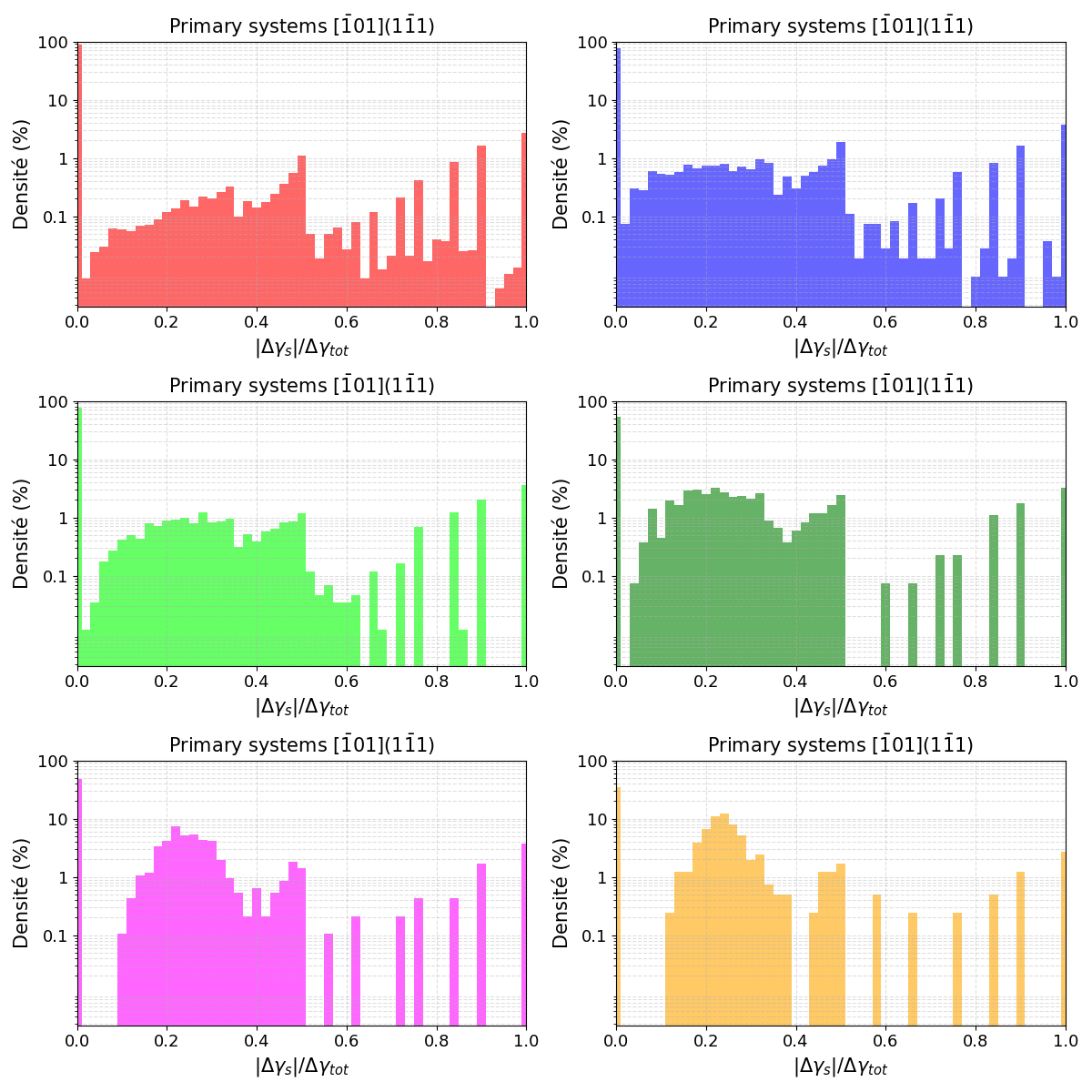}
\end{minipage}
\hfill
\begin{minipage}{0.48\linewidth}
    \centering
    \includegraphics[width=\linewidth]{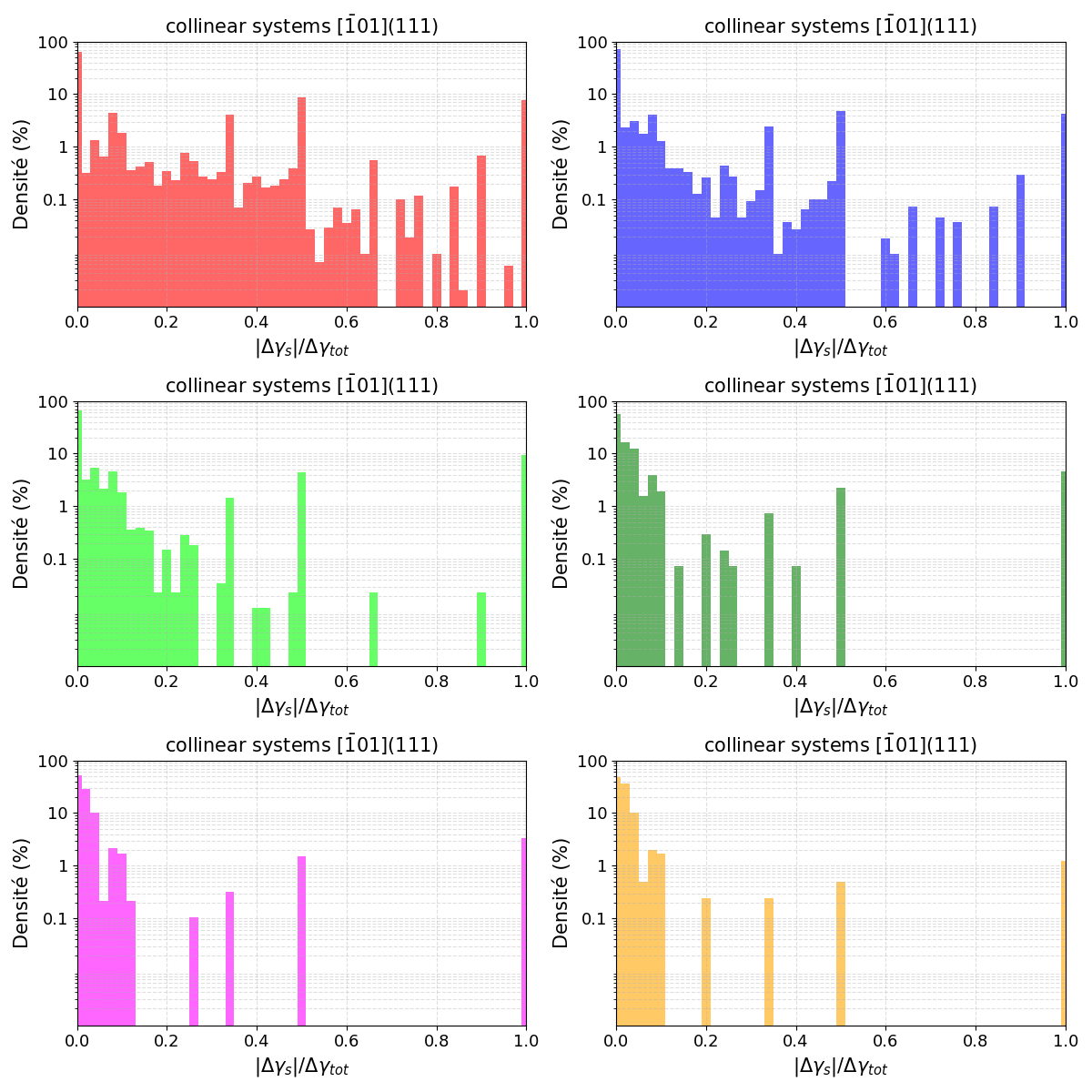}
\end{minipage}

\caption{{Relative contribution of individual slip systems 's' to dislocation 'a' (see main text) for different slip systems (columns). Each line correspond to a different imposed strain rate: 5/s (top row), 50 /s (middle row) and 500 /s (bottom row).}}
\label{contrib-slip syst}
\end{figure}

This section examines how the different slip systems contribute to dislocation avalanches. We propose a statistical analysis of the relative contributions of individual slip systems 's' though the ratio $c_{sa} = \frac{|\Delta \gamma_s| \in a}{\sum_j |\Delta \gamma_j| \in a}$ to an individualized avalanche 'a', with $\gamma_s$ the plastic strain provided by 's'. The absolute value sign is required to avoid compensation between contributions. This relative contribution is showed for the different slip systems contribute to dislocation avalanches. When $c_{sa}$ equals 100 \%, the corresponding slip system contributes fully to the considered avalanche, whereas when $c_{sa}$ equals 0, that system does not contribute at all. Histograms for the relative contribution $c_{sa}$ are provided for primary and collinear systems in Fig. \ref{contrib-slip syst} when analyzing all avalanches seen during one simulation. For the sake of brevity, we focus on the primary slip system $[\bar{1}01](1\bar{1}1)$ as well as on its associated collinear system $[\bar{1}01](111)$. The results for the remaining slip systems are presented in the supplementary material and exhibit the same overall features.

Histograms of $c_{sa}$ are markedly different for the various strain rates. 
At low strain rates, The histograms show a contribution spread across the entire range of $c_{sa}$ values, with a broad peak for  $0 < c_{sa} < 0.5$. In other words, at low strain rate, avalanches can be made of any relative contributions of slip systems. This is true for the primary and collinear systems contributions.  For larger strain rate however, the histograms become concentrated around 20-25\% for the primary systems and approximately 5-10\% for the collinear systems. In other words, avalanches all become very similar in terms of relative contributions with primary slip systems contributing almost equally to the avalanche, and a smaller contribution of collinear slip systems. In \cite{aissaoui2025physical}, we showed that CS as a slow rate process has to occur between avalanches.

The onset of an avalanche is associated to the destruction of a junction \cite{devincre2008dislocation}. Junction may be ternary or higher-order reactions, however most of the microstructure is made of binary junctions \cite{bulatov2006dislocation}. In order to get all slip systems contributing to dislocation avalanches at larger strain rate, it becomes obvious that several configurations evolving different subsets of slip systems must therefore be simultaneously activated. This, along with the larger activation stresses, show that more sites are activated simultaneously and the part of the dislocation density contributing to the critical part of plastic deformation increases [Aissaoui future paper on experiemental results]. Next we show the consequences on the avalanches properties.


\subsection{Size and duration of avalanches}\label{influnce-strain-rate-on-stat-avalanche}


\begin{figure}[h]
    \centering
    \subfloat[]{
        \includegraphics[width=0.45\textwidth]{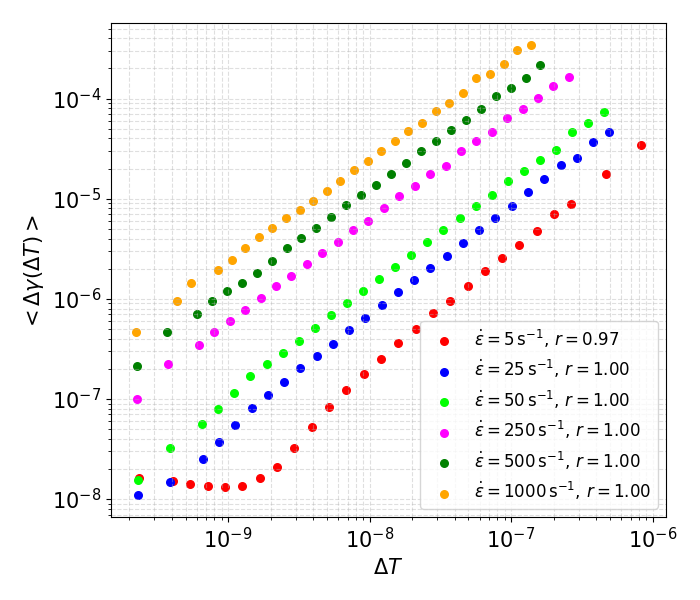}
    }
    \hfill
    \subfloat[]{
        \includegraphics[width=0.45\textwidth]{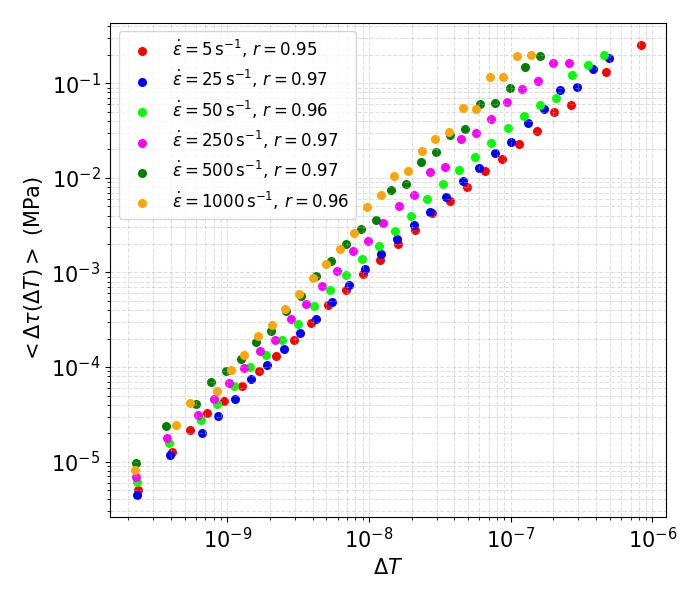}
    }
    \caption{Correlation of avalanche strain burst $\Delta \gamma$ (a) and stress drop $\Delta \sigma$ (b) with the avalanche duration $T$ for the different strain rate considered. 
    }
    \label{correlation-dt-dgamma-dtau} 
\end{figure}

In the remaining of the paper we draw the link between imposed strain rate and avalanche properties. {Figure \ref{correlation-dt-dgamma-dtau}(a) shows the correlation between strain-burst amplitudes ($\Delta\gamma$) and avalanche durations ($T$) for different strain rates. 
First, we observe that, for each strain rate, the correlation follows a behavior of the form $\Delta\gamma =C(\dot\epsilon) T^{r}$ for most of the duration range, where the exponent $r$ is indicated in the figure. All $r$ values are identified as close to 1, indicating a strong linear correlation between the strain-burst amplitude and the avalanche duration.  Furthermore, this correlation depends on the applied strain rate. As the strain rate increases from 5 to 1000 /s, for a given avalanche duration, the strain-burst amplitude also increases. Conversely, for a given avalanche size, the corresponding duration decreases with increased strain rate. In other words, avalanches become larger in size but shorter in duration as strain rate increases. This may be explained in different ways. i) Since multiple sites are activated simultaneously, the avalanche size $\Delta\gamma$ requires a smaller displacement from each activated configurations. ii) Larger applied stresses seen at larger strain rates, may yield to faster dislocation displacements, following the viscous mobility law employed. iii) at larger strain rates, activated configurations may be more out of mechanical balance past the activation stress. At a low strain rate of 5 and for avalanche durations below $10^{-9}$ s, the strain-burst amplitude remains nearly constant. Finally, it can be observed that the maximum avalanche duration increases while the maximum avalanche size decreases as the strain rate decreases. 


\begin{figure}[h]
    \centering

    \begin{subfigure}[b]{0.45\textwidth}
        \centering
        \includegraphics[width=\textwidth]{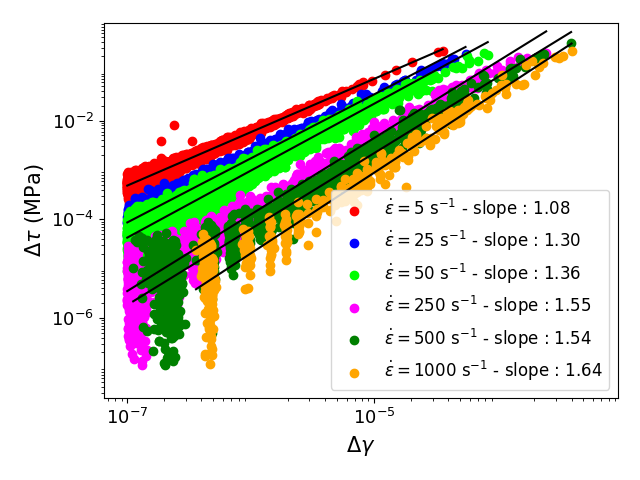}
        \caption{}
    \end{subfigure}
    \hfill
    \begin{subfigure}[b]{0.45\textwidth}
        \centering
        \includegraphics[width=\linewidth]{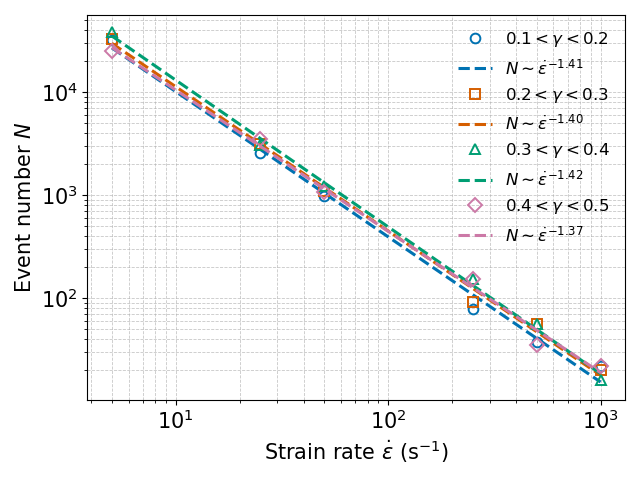}
        \caption{}
    \end{subfigure}

        \caption{a) Correlation between stress drop $\Delta \sigma$ strain burst $\Delta \gamma$ of avalanches as function of strain rate. b) Number of detected avalanche like events over a window of 0.1 \% of deformation and for various amount of deformation. 
        }
\label{fig:event_number}
\end{figure}

Let us now turn to Figure \ref{correlation-dt-dgamma-dtau}(b), which shows the correlation between stress drops ($\Delta\tau$) and avalanche durations ($T$) for the different strain rates, which shares a number of features with the previous analysis. For a given avalanche duration, the stress drop increases as strain rate increases.
This correlation also follows a power-law, $\Delta\tau \propto T^{\alpha}$, with $\alpha$ values ranging from 0.95 to 0.97. Interestingly, the stress drop seems to show a weaker dependence on the strain rate with curves for various strain rate rather close to each others.}

Since both $\Delta \sigma$ and $\Delta \gamma$ are both correlated to the avalanche duration $T$, these quantities are themselves correlated as shown in figure \ref{fig:event_number}. A large fluctuation can be seen on the data, which decreases with the event size. Moreover, the statistical fluctuations observed in the correlation curves tend to diminish as the system approaches a quasi-static regime for low strain rates. An average correlation of the form $\Delta\tau = A(\dot{\epsilon}) \cdot \Delta\gamma^{c}$ can be identified. This correlation appears to be affected by the evolution of the strain rate, with values of $c$ increasing as the strain rate increases. These correlations indicate that a given stress drop $\Delta\tau$ leads to larger strain bursts $\Delta\gamma$ at higher strain rates. Conversely, a given strain burst $\Delta \gamma$ will lead to a bigger stress drop $\Delta \tau$ as strain rate increases. This correlation is certainly connected to the loading control employed here.  For lower strain rates, the correlation exponent $c$ seems to tend to unity: $c \to 1$.  

Let us recall that the loading control employed here follows
$\mathrm{d}\sigma = E\bigl(\dot{\varepsilon}_a - \langle \dot{\varepsilon}_p \rangle\bigr)\,\mathrm{d}t$. where the change in the applied stress $d\sigma$ is linearly related to plastic strain increment $d\epsilon_p$. For a strain rate of 5, plastic burst $\Delta \gamma$ seen during individual avalanches are close to a linear correlation with stress drop  $\Delta \sigma$ with $c = 1.05$, which make the dislocation behavior directly related to the loading conditions, {which is consistent with what we expect from the quasistatic response of materials. This will be confirmed with the nice agreement seen on the exponent with experimental data \cite{weiss2015mild} .}

With more simultaneously activated sites and larger plastic events as strain rate increases, it is expected that the number of plastic events required to produce the same amount of strain has to decrease. This is confirmed in figure \ref{fig:event_number}.b) by the decreasing number of plastic events detected within a fixed window of (0.1\%). This explains the fewer serrations seen on the figure \ref{fig:epsi-point1}. The number of events follows as power law relation as $N \propto \dot {\epsilon}^{-1.4}$ independent of the amount of deformation reached, and this is not fully understood by the authors. A similar power law correlation was also observed on deformed Nb and Au pillars \cite{sparks2019avalanche}, however the exponent was found to be significantly different and this is certainly due to the finite size of the pillar systems that competes with the typical dislocation lengthscale. 

\subsection{Avalanche statistics}

\begin{figure}[h!]
    \centering
    \begin{subfigure}[b]{0.45\textwidth}
        \centering
        \includegraphics[width=\textwidth]{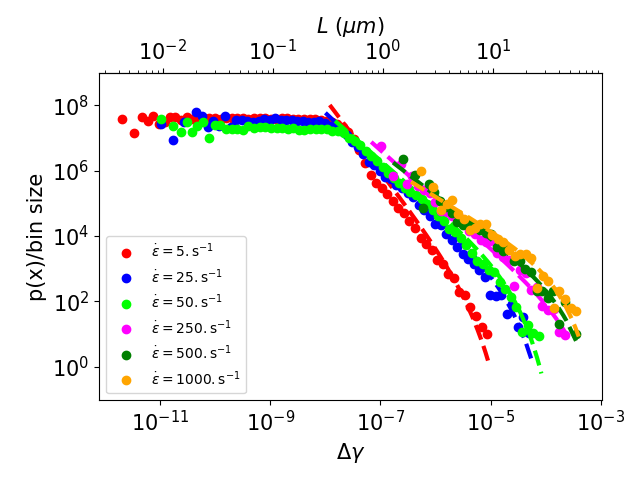}
        \caption{}
    \end{subfigure}
    \hfill
    \begin{subfigure}[b]{0.45\textwidth}
        \centering
        \includegraphics[width=\textwidth]{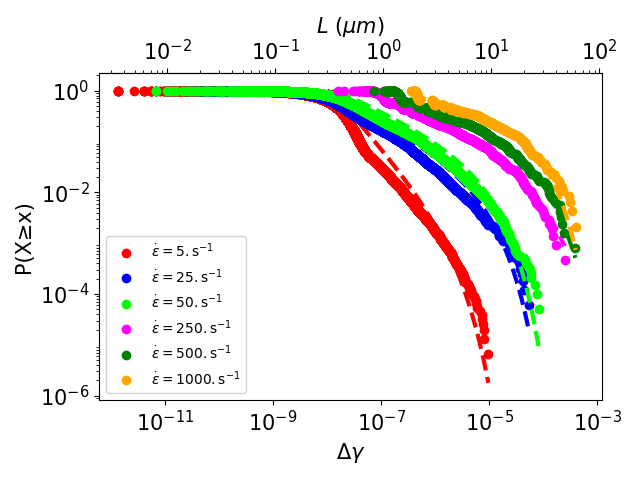}
        \caption{}
    \end{subfigure}
    \vfill
    \begin{subfigure}[b]{0.45\textwidth}
        \centering
        \includegraphics[width=\textwidth]{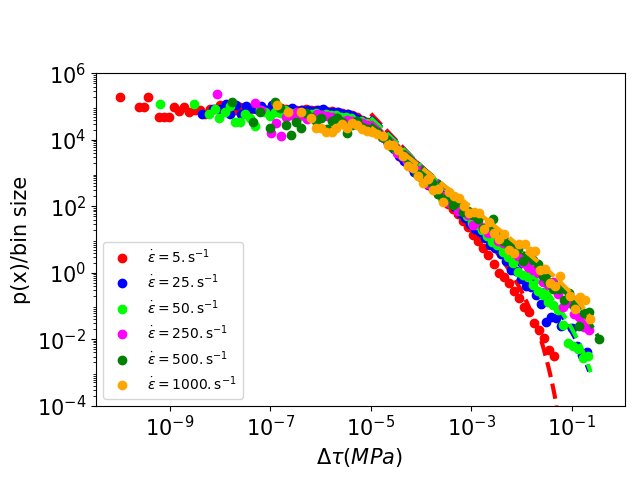}
        \caption{}
    \end{subfigure}
    \hfill
    \begin{subfigure}[b]{0.45\textwidth}
        \centering
        \includegraphics[width=\textwidth]{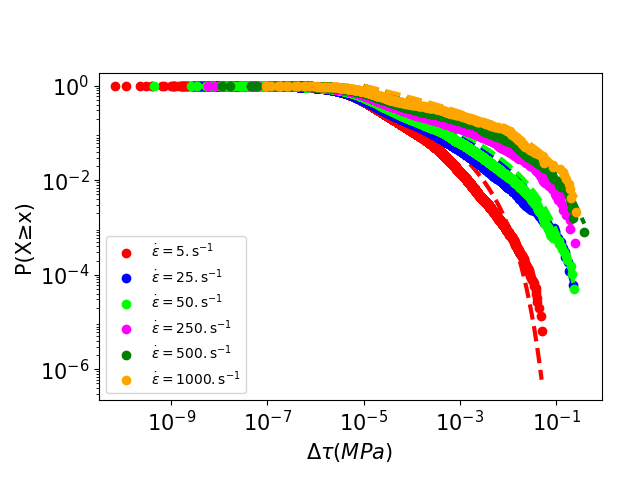}
        \caption{}
    \end{subfigure}
    \caption{Impact of applied strain rate on the statistical signature of dislocation avalanches in DDD simulations. (a) Probability density function (PDF) of the amplitudes of  strain bursts \(\Delta\gamma\) and the dislocation mean free path \(L~(\mu m)\). (b) Probability density function (PDF) of the stress drop amplitudes \(\Delta\tau\). (c) Complementary cumulative distribution function (CCDF) of the  strain bursts \(\Delta\gamma\) and the dislocation mean free path. (d) Complementary cumulative distribution function (CCDF) of the stress drops \(\Delta\tau\) for the three strain rates \(\dot{\epsilon}\). The simulations were carried out allowing cross-slip.}
\label{fig:avalanches_stats}
\end{figure}

The statistical signatures of dislocation avalanches are given in figure ~\ref{fig:avalanches_stats}, in terms of stress drops $\Delta\tau$ and strain bursts $\Delta\gamma$. In the following, we confirm that the distributions of those stress drops and strain bursts due to avalanches in FCC materials follow a power law with exponents close to those reported in the literature. In our previous study \cite{aissaoui2025physical}, we showed that the interaction types and dislocation density did not impact the power law exponent or lower cut-offs, but dramatically impacted the upper cut-offs modifying normalization constants. Here, we show that avalanches becomes larger and that the strain rate impacts all features of the avalanche statistics. 



Probability density functions in figure \ref{fig:epsi-point1}.a) and c) typically highlight the power law regime of avalanche statistics, and here this behavior is clearly visible for between 2 to 4 orders of magnitude in terms of $\Delta \gamma$ (and slightly more in terms of $\Delta \sigma$) depending on the strain rate. The exponent of the power law, quantified a bit further, decreases with strain rate. The power law regimes are bounded by well-defined cut-offs. The low cut-offs $\Delta \gamma_{min}$ is found to increase from about $\approx 10^{-8}$ to $\approx 10^{-6}$ as strain rate increases reducing the extent of the power law regime. To give an idea of the typical size of the strain bursts \( \Delta\gamma \) in our simulations, we define a length \( L \) analogous to the free path of an individual avalanche (whose average should converge towards the well-known mean free path, following the idea of Kubin~\cite{devincre2008dislocation}). Using Orowan’s relation \( \delta \gamma = \frac{L^2 b}{\text{Vol}} \), we can write: $L = \sqrt{\frac{\Delta\gamma \cdot \text{Vol}}{b}} $ where \( \text{Vol} \) is the simulation volume and \( b \) is the magnitude of the Burgers vector. $\Delta \gamma$ is also expressed in units of $L$ in the figures.

Power law upper cut-offs are typically better seen in complementary cumulative distribution function (CCDF) provided in figure \ref{fig:epsi-point1}.b) and d), where we see that they dramatically increase with the strain rate. The largest avalanches simulated are for example of about $\Delta \gamma_{max} \approx 10^{-5}$ ($L \approx 10 \mu$m) for a strain rate of 5 /s to $\Delta\gamma_{max} \approx 3.10^{-4}$ ($L \approx 70 \mu$m) for a strain rate of 1000 /s. a couple of remarks can be made. 
\begin{itemize}
    \item In our previous study, we showed that the upper cutoff decreased with the increase of dislocation density, following a deterministic scaling of the form $\Delta\gamma_{max} \propto b \sqrt{\rho}$. Here, simulations associated to large strain rates are also associated to larger dislocation densities, therefore the larger upper cut-offs are observed in spite of a larger dislocation densities.
    \item Besides, larger events seen at large strain rates are caused in part by the simultaneous activation of multiple sites. But, we argue that another cause to larger avalanches is the larger flow stresses applied. Stronger obstacles are certainly required to stop avalanches than in the case of low strain rates. 
\end{itemize}

Devincre and collaborators  \cite{devincre2008dislocation} showed that the dislocation evolution was related to plastic burst, as dislocation storage occurred at the mesoscale when avalanches come to a stop. Line length increase $\propto d\rho$ is related to the perimeter bounding the area $\propto d\gamma$ swept by dislocations \cite{queyreau2023saturation}. Therefore, larger avalanches observed at larger strain rate contribute to explain the increase in dislocation storage seen figure \ref{fig:epsi-point1}.b). This highlight the mutual interaction between strain rate sensitive behavior of materials and avalanche statistical signatures. 

To provide a more quantitative picture of impact of strain rate, we model the avalanche distributions of strain bursts using a power law truncated by an exponential decay of the form $P(\Delta\gamma) = A (\Delta\gamma)^{-\alpha_\gamma} \exp\left(-\frac{\Delta\gamma}{\Delta\gamma_{\text{max}}}\right)$ in the following, the integration of \( p(x) \) yields the cumulative distribution function \( P(x) \). We will use the same model for stress drops. The parameters of the probability distribution are estimated using the Maximum Likelihood Estimation (MLE) method proposed by Clauset~\cite{clauset2009power} and extended by Alstott~\cite{alstott2014powerlaw} in a Python implementation.

From the MLE analysis and the corresponding truncated power laws, we compute \( P(x) \) by integrating \( p(x) \), obtaining: $P(x) \propto \Gamma(1 - \alpha, \lambda x)$ where \( \Gamma \) is the upper incomplete gamma function and \( \lambda = \frac{1}{\Delta\gamma_{\text{max}}} \) is a non-universal cutoff parameter of the function \( f(x) \). From now on, $\Delta\gamma_{max}$ is a parameter modeling the upper truncation behavior of the avalanche statistics.  Our fits allow us to automatically estimate the scaling parameter \( \alpha \).


\begin{table}[ht]
\centering
\begin{tabular}{c|c|c|c|c|c|c}
\hline
\(\dot{\epsilon}\) (s\(^{-1}\)) & 5 & 25 & 50 & 250 & 500 & 1000 \\
\hline
\(\alpha_{\Delta\gamma}\)   & 2.25    &1.63     & 1.56    & 1.51    &   1.43  &  1.40    \\
\hline
\(1/\lambda_{\Delta\gamma}\)  & 2.65 $10^{-6}$    &   1.44 $10^{-5}$  & 1.63  $10^{-5}$  &   9.20 $10^{-5}$ &  1.32 $10^{-4}$   & 1.42  $10^{-4}$   \\
\hline
\(\alpha_{\Delta\tau}\)   &  1.67   &  1.52   &  1.50   &  1.39   & 1.35    &    1.31  \\
\hline
\(1/\lambda_{\Delta\tau}\)   &  7.38 $10^{-3}$  & 7.74 $10^{-2}$   &   7.76 $10^{-2}$ & 2.03 $10^{-1}$    & 2.04 $10^{-1}$   &   2.14 $10^{-1}$  \\
\hline
\end{tabular}
\caption{Quantitative modeling of avalanche statistics as function of the imposed strain rate for strain resolved (subscript $\gamma$) and stress resolved (subscript $\tau$) statistics.}
\label{tab:pwl_fit_values}
\end{table}

Table \ref{tab:pwl_fit_values} gathers the model parameters obtained by the MLE fitting. The parameters evolve monotonicaly with the strain rate and values will be discussed in the context of existing literature in next section.

\subsection{Discussion}


First, we showed that the strain rate sensitivity of fcc materials was associated to the activation and mobilization of dislocation configurations of increasing strength as strain rate increases. Weak configurations that are not producing enough plastic strain are easily destroyed and not even seen as a critical configuration. This is associated to a full microstructure organization as dislocation segments become smaller ($\propto 1/\sqrt{\bar a \rho}$) pinned by stronger obstacles of strength measured by $\bar a$. This original mesoscopic picture differs quite a bit from the strain rate dependence of Frank-Read sources of static properties typically considered until now.

This organization of the microstructure is transient by nature as seen in the evolution of the interaction strength in figure \ref{fig:beta-coefficient}. Takeuchi performed strain rate jump experiments on single crystals of Cu \cite{takeuchi1976variation}. Depending on the strain rate jumps and crystal orientation, transient stress peaks could be seen on the subsequent deformation curves and microstructure rearrangements were suggested. These observations match rather well our interpretation, with strong configurations formed at large strain rates being required to be destroyed before the microstructure evolve into a softer configuration, and was confirmed by some preliminary results. In \cite{edington1969influence}, TEM observations highlighted a larger dislocation density for larger strain and that forest mechanisms were suggested as mostly unchanged, which is in agreement with our simulation results. From initially single slip, the microstructures progressively included more activated systems as strain rate increases and could be explained by the need to activate more sites of the microstructure seen here. Finally, more edge dislocations are found in walls, as strain rate increases, microstructure contain more screw component. This is in agreement with our finding that collinear slip activity decreases with strain rate, and less time is left for cross slip activity between avalanches.  

\begin{figure}[h!]
\centering

\includegraphics[width=\linewidth]{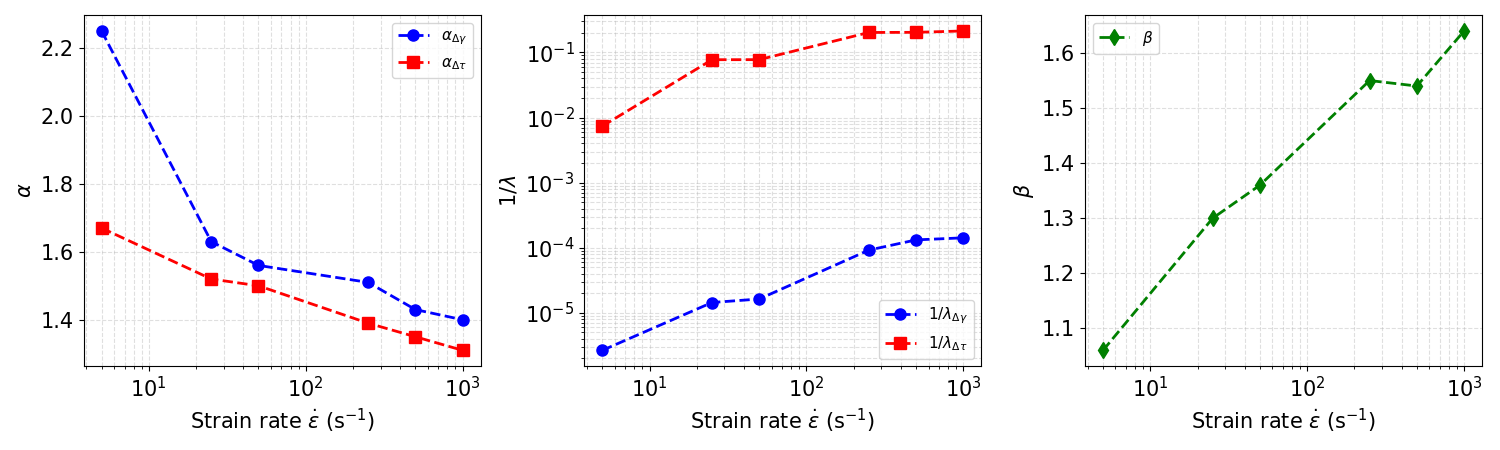}
\caption{Quantitative parameters defining the avalanche statistics as function of strain rate: a) exponents for the power law regime. b) Upper-cuttofs $1/\lambda$ and c) correlation exponent for $\Delta \sigma \propto \Delta \gamma^\beta$.}
\label{fig:param-distrib}
\end{figure}

Next, we can confront our results on avalanche statistics with other studies obtained in 3D plasticity. Specifically, the exponent $\alpha_{\gamma}$ decreases from 1.4 to 2.25 as strain rate decreases from 1000 to 5 /s as shown in figure \ref{fig:param-distrib}. For our lowest strain rate considered here of 5 /s, the flow stress $\tau_y$ and distribution of $p(\tau_{act})$ are not expected to evolve much as the strain rate sensitivity parameter $1/m$ is large. Also, the correlation exponent $r$ in $\Delta \tau \propto \Delta \gamma^r$ decreases from 1.64 to 1.08, and it can be argued that $r \to 1$ as strain rate decreases in agreement with the linear control employed here. Therefore, our simulation at 5 /s is close to this linear correlation, and $\alpha_{\gamma}$ is not expected to vary much as strain rate is decreased. Acoustic emission analysis \cite{weiss2015mild} lead to an amplitude exponent of $\alpha = 2.08 \pm0.15$ in Cu single crystals deformed as slow as $\dot\epsilon = 1.9\ 10^{-2}/s$, corresponding to the Arrhenius regime of strain rate dependence of plastic flow. The exponent for the lowest strain rate considered here may thus be seen as in a close agreement with the macroscopic experiments \cite{weiss2015mild,aissa2025}.

The change of exponent as function of loading rate has been reported on experimentally deformed micropillars \cite{sparks2018nontrivial, friedman2012statistics}, and was confirmed in our own macroscale experiments on Cu polycrystals \cite{aissa2025}. However, the quantitative comparison with the former is however delicate as the true strain rate is not easily accessible, and the finite size of samples competes with typical dislocation length-scale, hence modifying the power law cut-off behavior and reducing the extent of the power law regime. Nonetheless, in \cite{sparks2018nontrivial} the exponent of slip sizes was found to decreases from $\alpha = 2.02$ to 1.53, when the displacement rate increases from 6 to 600 nm/s on Au single crystals compressed along $<111>$. In another study, the size exponent was found to decrease from $\alpha = 2.1$ to $1.2$ as displacement rate increases from 0.1 to 10 nm/s on Au pillars \cite{friedman2012statistics}.

The relationship between the critical exponent and the strain rate has been theoretically discussed by White~\cite{white2003driving} in the context of "cracking noise," which invokes increased overlap of avalanches at higher loading rates. Furthermore, as the strain rate increases, avalanches become faster, and the distributions lose an increasing fraction of small events, shifting toward larger event sizes. These mechanisms reduce the value of the critical exponent. The dependence of the critical exponent on the strain rate indicates that plastic events do not belong to the same universality class.

Since the avalanche statistics in both $\Delta \gamma$ and $\Delta \tau$ are both identified as truncated power laws, $\Delta \tau$ and $\Delta \gamma$ are correlated in the average sense. Therefore, exponents are also correlated as: $-\alpha_{\tau} = {-\alpha_\gamma -r +1 \over r}$ and this matches rather well the values of $\alpha_{\tau}$ seen in figure \ref{fig:param-distrib}.a).

When now turning towards the largest strain rate considered here, comparison can be made with the work from Kuzunski and collab. who performed DDD simulations for rather large strain rate of 1000 /s and up \cite{kurunczi2023avalanches}. The authors obtained identical exponents $\alpha_\tau = \alpha_\gamma =1$ for analysis of event sizes in terms of $\Delta \gamma$ and $\Delta \tau$. This appears  at odds with our findings upon first consideration of the only strain rate common to both studies ($\dot\epsilon$ = 1000 /s). However, figure \ref{fig:param-distrib}.a) seems to show that both exponents increase with strain rate and become closer to one another. One could extrapolate from our data that $\alpha_\tau \to \alpha_\gamma \to 1$ as strain rate is increased.

The size of large events is estimated by the parameter $\lambda^{-1}$. This parameter depends on the initial dislocation density and the orientation. It corresponds to $\Delta\gamma_\mathrm{max}$ or $\Delta\tau_\mathrm{max}$ depending on the variable used to characterize the avalanche distributions. A strong dependence of $\lambda^{-1}$ on the strain rate is observed. Large events occur with higher probability as the strain rate increases, and the distributions of stress drops or strain bursts shift to the right, as shown in Figure ~\ref{fig:param-distrib}.


{Figure \ref{fig:param-distrib} presents the different power-law distribution parameters as a function of the strain rate. Figure \ref{fig:param-distrib}(a) shows the evolution of the critical exponent of the power-law distributions for stress drops and strain bursts. As discussed previously, a pronounced decrease of the critical exponent is observed at low strain rates. As the strain rate increases, the critical exponent progressively tends toward a constant value. A similar behavior is observed for the evolution of the power-law cut-off parameter \ref{fig:param-distrib}(b) : a strong variation at low strain rates, followed by a convergence of the cut-off values toward a unique value as the strain rate increases}



\newpage
\section*{Supplementary Material}

\subsection{Forest hardening vs strain rate hardening regimes}

\begin{figure}[h]
    \centering
    \subfloat[]{
        \includegraphics[width=0.45\textwidth]{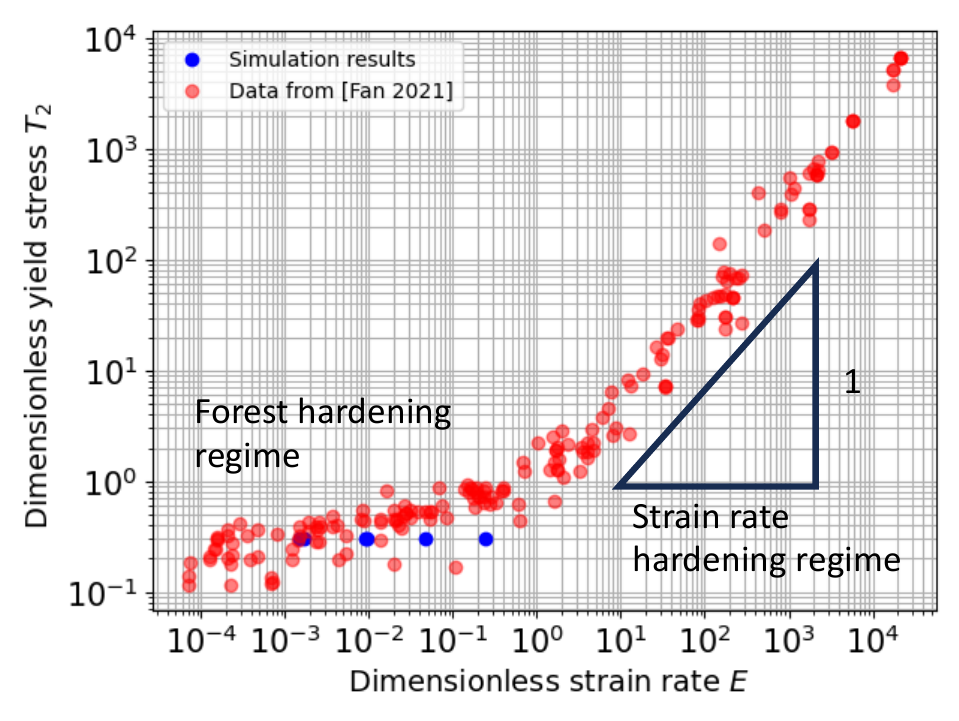}
    }
    \hfill
    \subfloat[]{
        \includegraphics[width=0.45\textwidth]{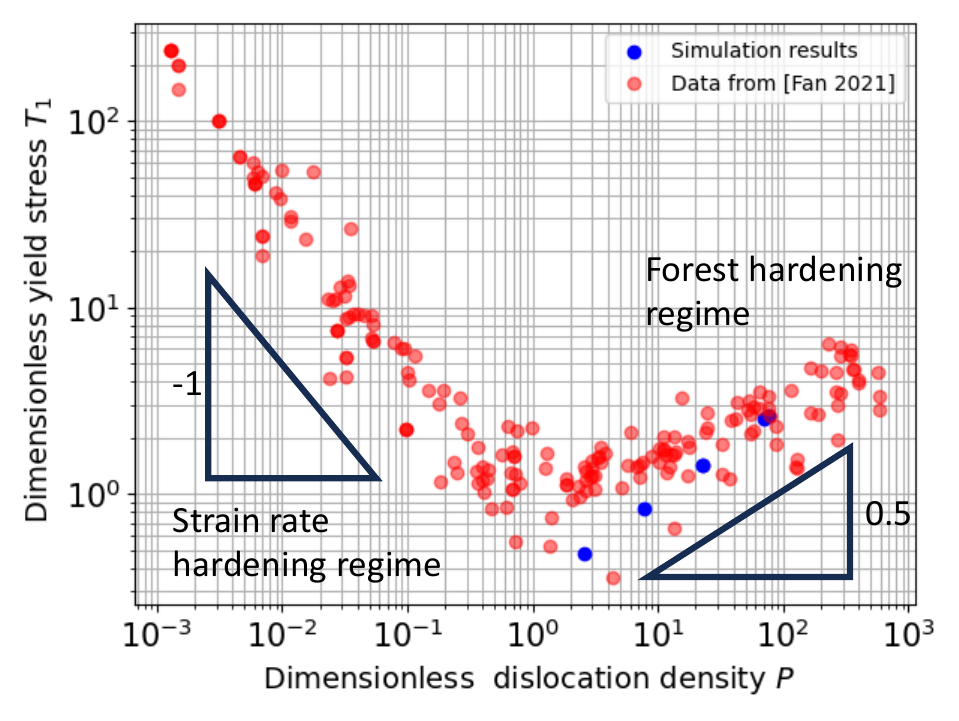}
    }
    \caption{Dimensionless parameters defining the forest hardening and strain rate hardening regimes. Dimensionless yield stress as function of a) dimensionless strain rate and b) dimensionless dislocation density. Some of our simulations are shown along with simulation and experimental data reviewed in \cite{fan2021strain}. }
    \label{fig:fan_parameters} 
\end{figure}

In \cite{fan2021strain}, plastic behavior was separated into two distinct regimes where i) plastic flow was controlled by forest hardening and ii) and a strain rate hardening regime where phonon dragging was the rate controlling mechanism. Since many materials or simulations parameters may impact the plastic behavior, dimensionless parameters for yield stress $T_1$ and $T_2$, dislocation density $P$ and strain rate $E$ were introduced and a first order model was proposed to model the two distinct regimes. Corresponding dimensionless parameters are calculated from our simulation conditions and showed in figure \ref{fig:fan_parameters}. For most of strain rate considered here, parameters belong clearly to the Forest hardening regime with yield stress $T_2$ weekly impacted by strain rate $E$  and $T_1 \propto \sqrt{P}$. The largest strain rate is just below the transition into the strain rate hardening regime.


\subsection{Behaviour when cross slip is disabled}

This section focuses on studying the influence of the strain rate on dislocation avalanches in the absence of cross-slip mechanisms. Although cross-slip plays a key role in crystalline plasticity, it is excluded here in order to isolate its direct effects of strain rate. Cross-slip is particularly sensitive to variations in strain rate since this mechanism is thermally activated. 


In Discrete Dislocation Dynamics (DDD), dislocations are assumed to be perfect and possess a well-defined slip plane. The possibility of cross-slip is incorporated by considering a probability for dislocations to deviate from one slip plane to another. Cross-slip in the FCC structure is modeled using the Escaig mechanism~\cite{escaig1968glissement}, based on a reaction pathway proposed by Friedel~\cite{friedel1957regarding}. This pathway consists of two steps. First, a local pinching of the dissociated dislocation in the primary plane, called a constriction, must occur. Then, the constriction dissociates into two half-constrictions, delimiting a segment of perfect dislocation, which immediately dissociates in the cross-slip plane.


\subsubsection{Plastic Behavior and Microstructure Evolution}

\begin{figure}[h]
    \centering
    \subfloat[]{
        \includegraphics[width=0.45\textwidth]{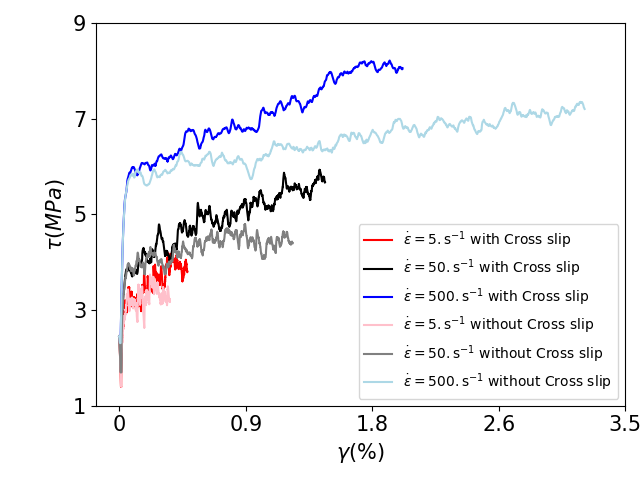}
    }
    \hfill
    \subfloat[]{
        \includegraphics[width=0.45\textwidth]{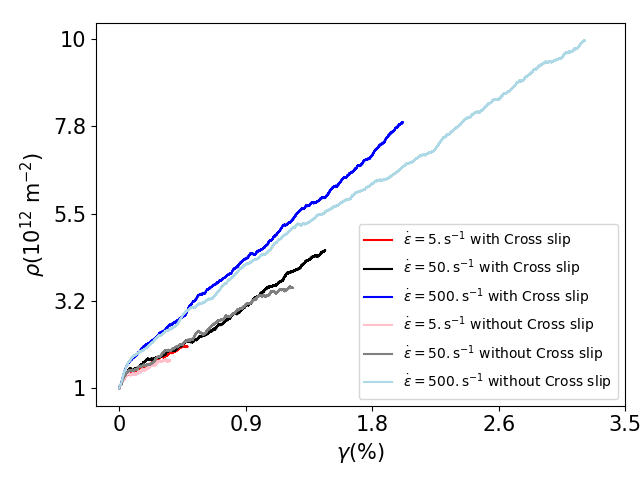}
    }
    \caption{Large-scale DDD simulations of [001] tensile deformation in Cu. The simulation box has a size of \(10 \, \mu\text{m}^3\), and the imposed strain rates are \(5, 50, 500 \ \text{s}^{-1}\) for two cases: one with cross-slip allowed and the other without cross-slip. (a) Corresponding evolutions of the resolved shear stress \(\tau\) as a function of the total shear strain \(\gamma\) for the three strain rates \(\dot{\epsilon}\). (b) Corresponding evolutions of the dislocation density \(\rho\) as a function of the strain \(\gamma\).}
    \label{fig:epsi-point3} 
\end{figure}

The simulations conducted to study the effect of strain rate in the absence of cross-slip were performed using the same parameters as those used in the previously presented simulations. This includes identical initial dislocation densities and simulation box dimensions. The resulting plastic behavior from these simulations is illustrated in Figure \ref{fig:epsi-point3}

The stress-strain curves shown in Figure \ref{fig:epsi-point3} (a) display, for the three strain rates, a decrease in the strain hardening rate when cross-slip is not activated. These results are confirmed by the strain hardening rate. This behavior is more pronounced for a strain rate of \(500\, s^{-1}\).

This influence of cross-slip has also been observed by Kubin et al. \cite{kubin1999dislocation} in the case of aluminum loaded along the high-symmetry $[001]$ axis. When cross-slip is taken into account, there is an increase in the number of collinear interactions. This promotes the formation of immobile debris and super-jogs, which hinder the motion of primary dislocations \cite{devincre2007collinear} and lead to additional hardening. In addition, the collinear interaction of jogs with other primary dislocations gliding on parallel planes contributes to this behavior \cite{devincre2006physical}.

Kubin et al. \cite{kubin1999dislocation} also reported that the dislocation microstructure in the presence of cross-slip shows a more pronounced organization. The annihilations and collinear super-jogs induced by cross-slip tend to anchor dislocations into the forming cell walls. The stress concentrations within these walls further promote cross-slip events. Queyreau \cite{queyreau2008etude} demonstrated that the activation of cross-slip during simulations involving a single active slip system leads to significant hardening, An increase in strain hardening associated with the activation of cross-slip has also been observed in DDD simulations of nanopillar compression \cite{hussein2015microstructurally}.

For strain rates of 5 and \(50\, s^{-1}\), the evolution of the dislocation density appears to be identical with or without cross-slip. However, simulations with a strain rate of \(500\, s^{-1}\) reveal a difference in the storage rate depending on the presence of cross-slip.

\subsubsection{Statistics of Dislocation Avalanches}
\begin{figure}[h!]
    \centering
    \begin{subfigure}[b]{0.45\textwidth}
        \centering
        \includegraphics[width=\textwidth]{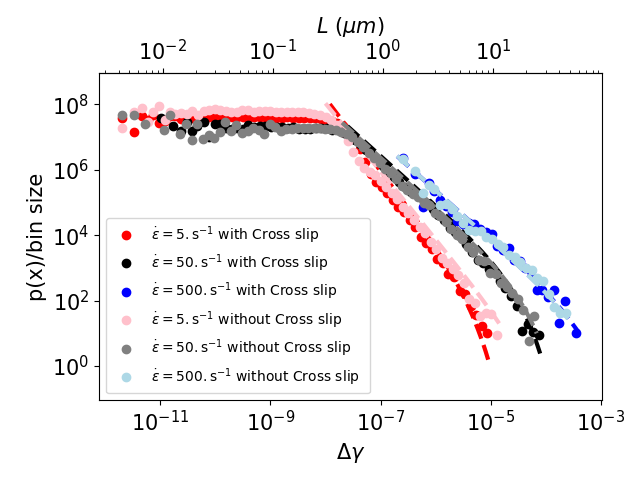}
        \caption{}
    \end{subfigure}
    \hfill
    \begin{subfigure}[b]{0.45\textwidth}
        \centering
        \includegraphics[width=\textwidth]{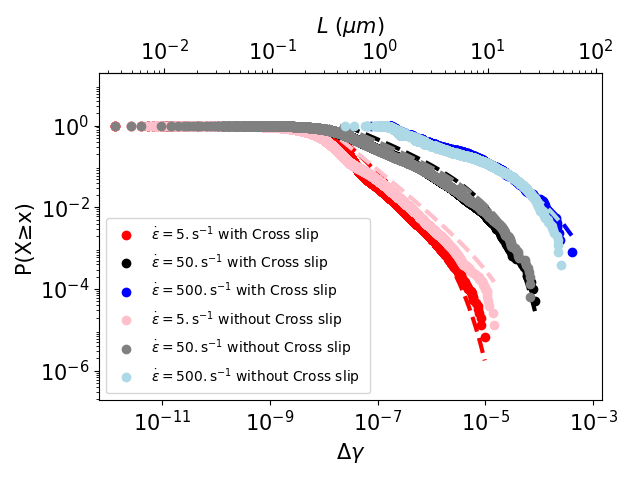}
        \caption{}
    \end{subfigure}
    \vfill
    \begin{subfigure}[b]{0.45\textwidth}
        \centering
        \includegraphics[width=\textwidth]{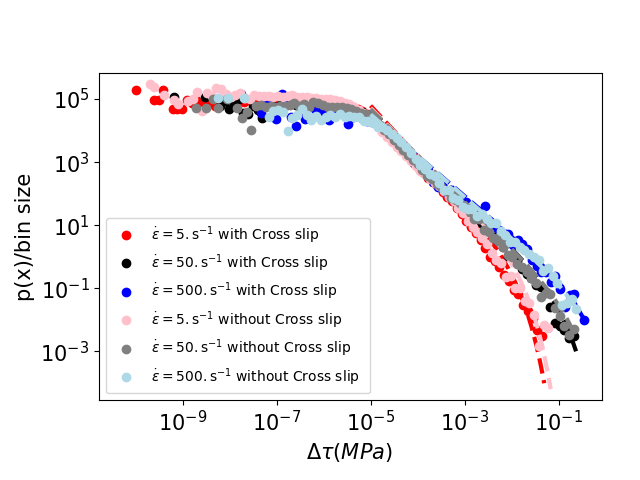}
        \caption{}
    \end{subfigure}
    \hfill
    \begin{subfigure}[b]{0.45\textwidth}
        \centering
        \includegraphics[width=\textwidth]{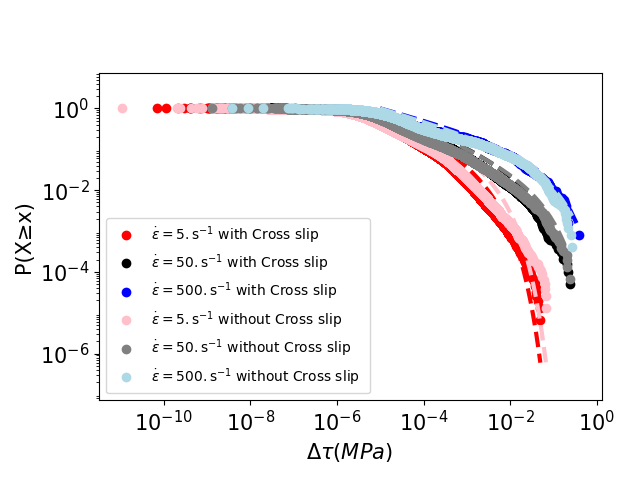}
        \caption{}
    \end{subfigure}
    \caption{Distribution of plastic events from a 3D DDD simulation at three imposed strain rates \(5, \ 50, \ 500 \ \text{s}^{-1}\) for two cases: one with cross-slip allowed and the other without cross-slip. (a) Probability density function (PDF) of the amplitudes of strain bursts \(\Delta\gamma\) and the dislocation mean free path \(L~(\mu m)\). (b) Probability density function (PDF) of the stress drop amplitudes \(\Delta\tau\). (c) Complementary cumulative distribution function (CCDF) of the strain bursts \(\Delta\gamma\) and the dislocation mean free path. (d) Complementary cumulative distribution function (CCDF) of the stress drop amplitudes \(\Delta\tau\) for the three strain rates.}
    \label{fig:epsi-point4}
\end{figure}

To better explain the influence of cross-slip on dislocation avalanches and its role during variations in strain rate, a statistical analysis of stress drops and  strain bursts was conducted for different strain rates, both with and without the possibility of cross-slip. The results are presented in Figures \ref{fig:epsi-point4}.

Figures \ref{fig:epsi-point4} (a) and \ref{fig:epsi-point4} (b) show, respectively, the probability density functions (PDFs) of  strain bursts $(\Delta\gamma)$ and stress drops $(\Delta\tau)$, while Figures \ref{fig:epsi-point4} (c) and \ref{fig:epsi-point4} (d) display the complementary cumulative distribution functions (CCDFs) of these same quantities. Since the simulations were performed in boxes of identical dimensions, the dislocation mean free path $L\,(\mu m)$ can be represented on the same graphs. The results exhibit trends similar to those presented in Section \ref{influnce-strain-rate-on-stat-avalanche}, namely that  strain bursts and stress drops follow truncated power-law distributions of the form $P(\Delta\tau) = A(\Delta\tau)^{-\alpha_\tau} \exp \left(- \frac{\Delta\tau}{\Delta\tau_0} \right)$ and $P(\Delta\gamma) = B(\Delta\gamma)^{-\alpha_\gamma} \exp \left(- \frac{\Delta\gamma}{\Delta\gamma_0} \right)$.

\begin{table}[ht]
    \centering
    \begin{tabular}{|c|c|c|c|}
        \hline
        $ \dot{\epsilon}~(\mathrm{s}^{-1}) $ & $5$ & $50$ & $500$ \\
        \hline
        \multicolumn{4}{|c|}{\rule[-2pt]{0pt}{15pt} \textbf{With cross-slip}} \\
        \hline
        $\alpha_{\Delta\gamma}$ & 2.25  & 1.65 & 1.45  \\
        \hline
        $\alpha_{\Delta\tau}$ & 1.67  & 1.50 &  1.35\\
        \hline
        $1/\lambda_{\Delta\gamma}$ &  2.65 $10^{-6}$ &  1.63 $10^{-5}$ &  1.53 $10^{-4}$ \\
        \hline
        $/\lambda_{\Delta\tau}$ &  7.38 $10^{-3}$ &  7.76 $10^{-2}$ &  2.04 $10^{-2}$  \\
        \hline
        \multicolumn{4}{|c|}{\rule[-2pt]{0pt}{15pt} \textbf{Without cross-slip}} \\
                \hline
        $\alpha_{\Delta\gamma}$ & 2.08  & 1.60 & 1.58  \\
        \hline
        $\alpha_{\Delta\tau}$ & 1.58 & 1.50 & 1.38 \\
        \hline
        $1/\lambda_{\Delta\gamma}$ &  3.32 $10^{-5}$ &  2.78 $10^{-5}$  &  3.64 $10^{-4}$ \\
        \hline
        $/\lambda_{\Delta\tau}$ &  9.24 $10^{-3}$  & 1.24 $10^{-1}$  &  2.41 $10^{-1}$ \\
        \hline
       
    \end{tabular}
    \caption{}
    \label{tab:valeur-wcs}
\end{table}

The results obtained in both cases—whether cross-slip is allowed or forbidden, are very similar. These distributions exhibit power-law behavior that depends solely on the strain rate. The critical exponents, as well as the characteristic parameters of the distributions, such as ($x_{min}$) (the minimum value from which the power law is defined) and the cut-offs (characteristic maximal values associated with large events), are influenced by the strain rate but are not affected by the activation or suppression of cross-slip.

The cross-slip mechanism is intrinsically slower than the dynamics of dislocation avalanches. Cross-slip events typically occur during stress build-up phases, while dislocation avalanches correspond to abrupt stress release events (stress drops). Cross-slip does not have a significant influence on the statistical distribution of avalanche characteristics $\Delta\gamma$ and $\Delta\tau$, which are primarily sensitive to strain rate. These results highlight the importance of loading conditions on avalanche dynamics and demonstrate that certain slower, local mechanisms have a limited impact on collective phenomena.

\newpage

\bibliographystyle{unsrt}
\bibliography{name}
\end{document}